\documentclass[aps,prd,amsmath,amssymb,preprintnumbers,onecolumn,11pt,nofootinbib]{revtex4}
\usepackage{mathtools} 
\usepackage[a4paper,left=20mm,right=20mm,top=25mm,bottom=25mm]{geometry}
\usepackage{mathrsfs}
\usepackage{graphicx,epsfig}
\usepackage{color}
\usepackage{fancyhdr}
\usepackage{multirow}
\usepackage{tikz}
\usepackage{feynmp}
\usepackage{mathpazo}
\usepackage{bbold}
\usepackage{natbib}
\usepackage{diagbox} 
\usepackage{comment}

\usepackage{amsfonts}
\usepackage{bm}
\usepackage{xcolor}

\definecolor{CP3}{cmyk}{0,0.88,0.77,0.40}

\newcommand\be{\begin{equation}}
\newcommand\ee{\end{equation}}
\newcommand\ba{\begin{eqnarray}}
\newcommand\ea{\end{eqnarray}}

\newcommand{\mpl}{m_p}
\newcommand{\tV}{\tilde{V}}
\newcommand{\tphi}{\tilde{\phi}}

\newcommand{\call}{{\cal L}}

\newcommand{\lo}{\lambda_0}

\renewcommand{\(}{\left(}
\renewcommand{\)}{\right)}
\renewcommand{\[}{\left[}
\renewcommand{\]}{\right]}

\newcommand{\N}{\mathcal{N}}

\newcommand\fc{f_{,c}}
\newcommand\fcc{f_{,cc}}

\newcommand\vp{V_{\phi}}

\newcommand{\cH}{\mathcal{H}}

\begin{document}

\title{\Large \color{CP3} Inflationary model in minimally modified gravity theories}
\author{Jakkrit Sangtawee $^{1,2}$}
\email{jakkrits60@nu.ac.th} 
\affiliation{\footnotesize $^{1}${The Institute for Fundamental Study \lq\lq The Tah Poe Academia Institute\rq\rq, \\Naresuan University, Phitsanulok 65000, Thailand}}
\author{Khamphee Karwan$^{1,2}$}
\email{khampheek@nu.ac.th}
\affiliation{\footnotesize $^{1}${The Institute for Fundamental Study \lq\lq The Tah Poe Academia Institute\rq\rq, \\Naresuan University, Phitsanulok 65000, Thailand}}
\affiliation{\footnotesize $^{2}${Thailand Center of Excellence in Physics, Ministry of Higher Education, Science, \\Research and Innovation, Bangkok 10400, Thailand}}

\begin{abstract}

We have investigated inflationary model constructed from minimally modified gravity (MMG) theories.
The MMG theory in the form of $f({\bf H}) \propto {\bf H}^{1+p}$ gravity where, ${\bf H}$ is the Hamiltonian constraint in the Einstein gravity and $p$ is constant, has been studied.
An inflation is difficult to be achieved in this theory of gravity unless an additional scalar field playing a role of inflaton is introduced in the model.
We have found that the inflaton with exponential potential can drive inflation with graceful exist different from the case of Einstein gravity.
The slow-roll parameter for both the exponential and the power-law potentials is inversely proportional to number of e-folding similar to the case of the Einstein gravity.
We also have found for the scalar perturbation that the curvature perturbation on super Hubble radius scales grows rapidly during inflation unless $p \sim 0$.
For the tensor modes, the amplitude of the perturbations is constant on large scales up to the lowest order in slow-roll parameter, and sound speed of the perturbations  can diviate from unity and can vary with time depending on the form of $f({\bf H})$.

\end{abstract}

\maketitle

\section{Introduction}

Cosmic inflation \cite{Guth:1980zm, Linde:1981mu, Albrecht:1982wi} is a standard framework addressing issues in the hot Big Bang model and providing mechanism for generation of primordial density perturbation. 
In the standard scenario, inflation can be achieved by introducing extra degrees of freedom in the universe. 
In the case of the Einstein gravity the extra degrees of freedom may be in the form of fields minimally couple to gravity called inflaton.
Alternatively, the extra degrees of freedom can be parts of degrees of freedom of the gravitational interaction.
The extra degrees of freedom of gravity can be obtained by assuming non-minimally coupling between extra field and curvature terms in the action.
This class of theories is scalar-tensor theories of gravity \cite{Fujii2009}.
Moreover, the extra degrees of freedom of the gravitational interaction can also be obtained due to non-linear curvature terms in the action. 
The simplest example of this class of gravity is $f(R)$ gravity \cite{Clifton:11}.

However, in the cuscuton models \cite{cuscut, Iyonaga:18, Iyonaga:20}, it has been shown that the acceleration of the universe can be achieved even though the minimally couple extra degree of freedom is non-dynamical  field. 
This implies that theories which have two dynamical degrees of freedom can also drive acceleration of the universe.
Theories of gravity beyond the Einstein theory with have two degrees of freedom as the Einstein theory have been studied in various contexts \cite{cov, Mukohyama:17, Katsuki:18,Aoki:2018brq, Mukohyama:19, DeFelice:2020eju, Gao:20, Yao:21, Hu:21}. 
Such theories could be constructed by supposing that the temporal diffeomorphism is broken while the spatial diffeomorphism is still invariant.
In general, if the diffeomorphism invariant is broken in this way, the theories can have an extra degree of freedom similar to scalar-tensor theories of gravity \cite{Gao:14}.
However, if the Lagrangians of theories are a linear function of the lapse function, 
the theories can have two tensorial degrees of freedom for gravity under suitable conditions.
This class of theories is minimally modified gravity theories \cite{Mukohyama:17, Mukohyama:19}. 
Nevertheless, these conditions can not be satisfied if matter appears in the action. 
To ensure that this class of theories still has two tensorial degrees of freedom for gravity when matter appears in the theories, we have to impose the gauge fixing condition \cite{Katsuki:18,Carballo-Rubio:2018czn,Aoki:20}.
Cosmology with this class of theories has been investigated in \cite{Iyonaga:20, Aoki:20}. 
In \cite{Aoki:20}, it has been shown that late time universe with this class of gravity theories is more preferred by observational data than $\Lambda$CDM model.
In \cite{Lin:18, Lin:19, Aoki:20}, matter coupling in this class of theories has been discussed.

Here, we investigate inflation due to this class of gravity theories.
This work is organized as follow: firstly we review MMG theories in the next section. 
 We investigate background inflation in Sec.~(\ref{sec:3}). 
We study cosmological perturbation in Sec.~(\ref{sec:4}) and conclude in the last section.  

\section{Minimally modified gravity theories}
\label{sec:2}

Minimally modified gravity theories  are the modified theories propagating two tensorial degrees of freedom like the Einstein theory of gravity. 
Generally, most of popular modified theories of gravity always generate
extra degrees of freedom in the theories. The extra degrees of freedom can be related to the broken 
diffeomorphism invariant in the construction of the theories. However, we can construct the theories
that have two tensorial degrees of freedom even if the full diffeomorphism invariant is broken. We can construct
MMG theories by supposing that the Hamiltonian of the theories is linear in the lapse function and imposing a suitable constraint. Square root 
gravity and exponential gravity are the MMG theories that we obtain by using this method \cite{Mukohyama:17}.
However, there is an interesting class of MMG theories, $f({\bf H})$ theory,
   in which Lagrangian of the theory is a function of the Hamiltonian constraint ${\bf H}$ in the Einstein gravity
This class of MMG theories is constructed in another way by the Hamiltonian construction \cite{Mukohyama:19}. 

In order to construct  the MMG theories, we break the temporal diffeomorphism invariant which is conveniently represented by the ADM decomposition. 
In the ADM formalism, one can write the line-element in the form
	\be \label{adm}
	ds^2 = \(-\N^2 + \N_i\N^i\)dt^2 + h_{ij}\(\N^i dt + \N dx^i\)\(\N^j dt + \N dx^j\)\,,
	\ee
where $h_{ij}$, $\N$ and $\N^{i}$ are the three-dimensional induce metric, the lapse function and the shift vector, respectively. 
We are interested in MMG theories in the form of $f({\bf H})$ theory which the action can be written in the form 
\be
\label{minimalLag}
S[h_{ij},\N,\N^i] = \frac{\mpl^2}2 \int d^4x \N \sqrt{h} \call_G
= \frac{\mpl^2}2 \int d^4x \N \sqrt{h} \left[ \frac{2}{\fc(C)}(K_{ij} K^{ij} - K^2) - f(C)  \right]\,,
\ee
where $\mpl = 1/\sqrt{8\pi G}$ is the reduced Planck mass, $h$ is the determinant of the metric $h_{ij}$ and
\be
K_{ij} = \frac{1}{2 \N} \( \dot{h}_{ij} - D_{j} \N_{i} - D_{i} \N_{j} \) \,.
\ee
Here, $D_i$ is the covariant derivative compatable with the metric $h_{ij}$ and a dot denotes derivative with respect to time $t$.
The variable $C$ can be computed from
\be
\label{constraintC0}
C = \frac{K_{ij} K^{ij} - K^2}{[\fc(C)]^2} - R \,.
\ee
From the above expressions, $f(C)$ is an arbitrary function of $C$, $\fc$ denotes derivative of $f(C)$ with respect to $C$, and we see that $C$ has the same dimension as the three-dimensional Ricci scalar $R$, i.e., its dimension is mass${}^2$.
Moreover, the action reduces to the action for the Einstein gravity if $\fc = 1$. 

To study possible models of inflation from this theory of gravity, we add extra scalar field into the above action as
\be
\label{action}
S[h_{ij},\N,\N^i, \phi] =  \int d^4x \N \sqrt{h} \left[ \frac{\mpl^2}{2} \call_G + X - V(\phi) \right] \,.
\ee
Here, we suppose that the field has standard kinetic term where $X = -\partial_{\mu} \phi \partial^{\mu}\phi/2$ is the kinetic term of the scalar field and $V$ is the potential term.
However, the degree of freedom in the theory increases when the scalar field is simply added in the action.
To ensure that the theory still has two tensorial degrees of freedom for gravity, we have to fix the gauge degree of freedom in the theory.
Using the choice of gauge presented in \cite{Aoki:20}, the Hamiltonian of the gauge fixeing term is written in the form
	\be
	H_{gf} = \int d^3x \sqrt{h} \tilde{\lambda}^{i} \partial_{i} \( \frac{\pi}{\sqrt{h}} \) \,,
	\ee 
where $\tilde{\lambda}^{i}$ is a Lagrange multiplier and $\pi$ is the trace of momentum conjugate to the induce metric.
Imposing this gauge fixing, the action for $f({\bf H})$ becomes
	\ba \label{action:ful}
	S &=& \frac{1}{2} \int d^4x \N \sqrt{h} \Bigg\{ \mpl^2 \bigg[(C+R)\[ 2 - \lambda_{0} \fc(C) \]\fc(C) - f(C) \nonumber\\
	&&+ \lambda_{0} \[ K^{ij}K_{ij} - K^{2} - \frac{2K}{\N} D_{k} \tilde{\lambda}^k - \frac{3}{2\N^{2}} \( D_{k} \tilde{\lambda}^k   \)^{2} \] \bigg] + 2X - 2V(\phi) \Bigg\} \,,
	\ea
where $\lambda_{0}$ is another Lagrange multiplier, and in this case $C$ becomes
	\be
\label{constraintC}
	C = \frac{1}{\[ \fc(C) \]^{2}} \[ K^{ij}K_{ij} - K^{2} -\frac{2K}{\N} D_{k} \tilde{\lambda}^k - \frac{3}{2\N^{2}} \( D_{k} \tilde{\lambda}^k   \)^{2} \] -R \,.
	\ee
The above expression for $C$ can be obtained by varying the action Eq.~(\ref{action:ful}) with respect to $\lo$.
Varying the action with respect to $C$, $\N$ and $\N^k$ yiels, respectively, 
\ba
\lo &=&
	\frac{1}{\fc}\,, \label{vary-l1}\\
0 &=& 
	f(C)- \frac 2{\mpl^2}\[X-V - \frac{1}{\N^2}\(\dot\phi - \N^i\partial_i\phi\)^2\]\,,	 \label{encon}  \\
0 &=&
	D_i K^{ik} - h^{ik} D_i K - h^{ik} D_i D_m \lambda^m_0 - \frac{1}{\mpl^2 \N} \( \dot{\phi} - \N^i \partial_i \phi \) \partial_k \phi \,.
\ea
Variation with respect to scalar field give us the evolution equation for scalar field as
\be
\frac{\partial}{\partial t}\[ \frac{\sqrt{h}}{\N} \( \dot{\phi} 
- \N^{j} \partial_{j} \phi \) \]
- \partial_{i} \[ \frac{\sqrt{h}}{\N} \N^{i} \dot{\phi} + \N \sqrt{h} \( h^{ij} - \frac{\N^{i} \N^{j}}{\N^2} \) \partial_{j} \phi  \] + \N \sqrt{h} V_{\phi}= 0\,,
\ee
where subscript ${}_\phi$ denotes derivative with respect to scalar field $\phi$.

\section{Background evolution}
\label{sec:3}

We now consider the evolution of the spatially flat Friedmann universe for the theory described in the previous section. 
Due to the homogeneity and isotropy of the Friedmann universe, $\N = 1$, $\N^i = 0$ and therefore
\be
ds^2 = - dt^2 + a^2(t)\delta_{ij}dx^i dx^j\,,
\label{flrwds2}
\ee
where $a(t)$ is a cosmic scale factor.
For the Friedmann universe, the constraint in Eq.~(\ref{encon}) and the expression for $C$ in Eq.~(\ref{constraintC}) are given by
\ba
f &=& - \frac{2}{\mpl^2} (X + V) = - \frac 1{\mpl^2} \(\dot\phi^2 + 2 V\)\,,
\label{enconv2}\\
C \fc^2 &=& - 6 H^2\,,
\label{ceqbk}
\ea
where $H \equiv \dot a / a$ is the Hubble parameter.
The evolution equation for scalar field inflaton in the Friedmann universe is
\be
\ddot\phi + 3 H \dot\phi + \vp = 0\,.
\label{kgbk}
\ee
The slow-roll parameter $\epsilon \equiv - \dot H / H^2$ can be computed by differentiating Eq.~(\ref{ceqbk}) with respect to time to obtain $\dot{C}$,
and substituting resulting $\dot{C}$ into the time-derivative of Eq.~(\ref{enconv2}).
The result is
\be
\epsilon = \frac{\eta \fc}{2} \(1+ 2 \frac{C\fcc}{\fc}\)\,,
\label{ep:gen}
\ee
where $\eta \equiv \dot\phi^2/(H^2 \mpl^2)$.
The above relation reduces to the usual relation for $\epsilon$ for the Einstein gravity when $\fc = 1$.
It follows from Eq.~(\ref{ep:gen}) that $\epsilon\ll 1$, which is required during inflation, when $\eta \ll 1$ or $|\fc + 2 C \fcc| \ll 1$.
However, the latter condition is difficult to be achieved, so that slow-roll inflaton is need for inflation in this theory.
The case $\eta \ll 1$ corresponds to the slow-roll evolution of the inflaton field $\phi$.
Under the slow-roll approximation, $|\ddot\phi| \ll |H \dot\phi|$,
Eq.~(\ref{kgbk}) becomes
 \be
\frac{d\phi}{dN} = - \frac{\vp}{3 H^2}\,,
\label{phi:slow}
\ee
where $N \equiv \ln a$ is the number of e-folding.

In order to study the evolution of the background universe,
we have to specify form of $f(C)$.
Here, we suppose
\be
f(C) = - \Lambda \( - \frac{C}{\Lambda}  \)^{1+p}\,,
\label{deff}
\ee
where $\Lambda$ is a constant with dimension of mass${}^2$ and $p$ is a constant parameter.
We then obtain   from Eq.~(\ref{ceqbk}) that
\ba
f &=& - \Lambda \[ \frac{6 H^2}{\Lambda (1+p)^2} \]^{ \frac{1+p}{2p+1} }\,,
\label{f:model}\\
\fc &=& (1+p) \[ \frac{6 H^2}{\Lambda (1+p)^2} \]^{ \frac{p}{2p+1} }\,.
\label{fc:model}
\ea
Hence, we obtain the modified Friedmann equation by substituting the above expression into Eq.~(\ref{enconv2}) as
	\be
	\[ \frac{6 H^2}{\Lambda (1+p)^2} \]^{\frac{1+p}{2p+1}} = \frac{1}{\mpl^2 \Lambda} \( \dot{\phi}^2 + 2 V(\phi) \)		\,.
	\label{mod:fd}
		\ee
Using slow-roll condition, $V \gg \dot{\phi}$,
we can write Eq.~(\ref{mod:fd}) as
	\be
	H^2 = \frac{ 2^{\frac{2p+1}{1+p}} (1+p)^2 \Lambda }{6} \( \frac{V}{\mpl^2 \Lambda} \)^{ \frac{2p+1}{1+p} }    \,.
	\label{11}
	\ee
Substituting Eq.~(\ref{11}) into Eq.~(\ref{phi:slow}) we get 
	\be
	\frac{d \phi}{d N} = -	\frac{2^{-p/(1+p)}}{\Lambda (1+p)^2 } \frac{ V_{\phi} }{\tV^{\frac{2p+1}{1+p}}}	\,,		\label{difeq}
	\ee
where $\tV \equiv V / (\mpl^2 \Lambda)$.
The above equation can be written in the integral form as
\be
	\int_0^{N_N} d N = 2^{p/(1+p)} \Lambda \( 1+p \)^2 \int_{\phi_e}^{\phi_N} d \phi \frac{\tV^{ \frac{2p+1}{1+p} }}{V_{\phi}}\,,		\label{inteq}
	\ee
where subscript ${}_e$ denotes evaluation at the end of inflation,
while subscript ${}_N$ denotes evaluation at the moment when particular modes of cosmological perturbations generated during inflation crosses the horizon.
For the form of $f$ given by Eq.~(\ref{deff}),
the slow-roll parameter $\epsilon$ in the slow-roll approximation is
	\be
	\epsilon = \frac{2^{-\frac{2p+1}{1+p}} \(2p+1\) }{\mpl^2 \Lambda^2 \(1+p\)^3} \frac{V_{\phi}^2}{\tV^{\frac{3p+2}{1+p}}}\,.
	\label{ep:model}
	\ee
In the slow-roll approximation,
we can write $\fc$ in terms of the potential as
\ba
\fc &=& (1+p) 2^{\frac{p}{1+p}} \tV^{ \frac{p}{1+p}}\,,
\label{fc:slow}\\
C &=& - \Lambda 2^{\frac{1}{1+p}} \tV^{ \frac{1}{1+p}}\,.
\ea 
To integrate Eq.~(\ref{inteq}), and compute $\epsilon$ in terms of the number of e-folding,
we have to specify the potential $V$ of scalar field.
As the illustrative examples,
we will consider two cases where $V$  takes either exponential or power-law form.

\subsection{Exponential potential}
\label{exp:bg}

We first consider the potential in the form
	\be
	V(\phi) = V_0 \Lambda \mpl^2 e^{\lambda \tphi}\,,
	\label{vexp}
	\ee
where $\tphi \equiv \phi/\mpl$, while $V_0$ and $\lambda$ are the dimensionless constants.
Substituting the above potential in Eq.~(\ref{inteq}),
and performing an integration,
we get
\be									\label{Nfp}
	N_N = \frac{ 2^{p/(1+p)} \(1+p\)^3  }{\lambda^2 p V_0^{-p/(1+p)}} \[e^{\lambda \tphi_N p/(1+p)} 
		- e^{\lambda \tphi_e p/(1+p)}  \]\,.
	\ee
We can calculate $\phi_e$ by using the slow-roll parameter.
Since $\epsilon = 1$ at the end of inflation,
we get from Eq.~(\ref{ep:model}) that
	\be \label{field-evolution-exponential}
	e^{\lambda \tphi_e p/(1+p)} = \frac{\lambda^2 (2p+1)}{ 2^{\frac{2p+1}{p+1}} (1+p)^3 V_0^{p/(1+p)}  }\,.
	\ee
Substituting the above equation into Eq.~(\ref{Nfp}), we get
	\be
	N_N + N_* = \frac{ 2^{p/(1+p)} (1+p)^3  }{ \lambda^2 p V_0^{ -p/(1+p) } }e^{\lambda \tphi_N p/(1+p)}\,,
\label{nn:exp}
\ee
where 
\be \label{n*:exponential}
N_* \equiv \frac{2p+1}{2 p}\,.
\ee
Inserting Eq.~(\ref{nn:exp}) into Eqs.~(\ref{ep:model}) and (\ref{phi:slow}),
we can write $\epsilon$ and $\eta$ in terms of the number of e-folding as
\be 
\epsilon_N = \frac{N_*}{N_N + N_*}\,,
\quad
\eta_N = \frac{(p+1)^2 }{\lambda^2 p^2 (N_N + N_*)^2}
\,.
\label{epneta:exp}
\ee
Using Eqs.~(\ref{fc:slow}) and (\ref{nn:exp}),
we have
\be 
\fc(N) = \fc{}_* (N_N + N_*)
= \frac{\lambda^2 (2p+1)}{2 (1+p)^2} \frac{1}{\epsilon}\,,
\label{fc:exp}
\ee
where $\fc{}_*$ is defined as
\be
\fc{}_* \equiv \frac{\lambda^2 p}{(1+p)^2}\,.
\ee
It follows from the above calculations that the inflaton with exponential potential has graceful exit in this theory of gravity.
This result is different from that in Einstein theory of gravity. 
The moment at graceful exit is described by Eq.~(\ref{field-evolution-exponential}). 

\subsection{Power-law potential}
\label{pwl:bg}

In this section, we apply the potential of the form,
	\be
	V(\phi) = V_0 \mpl^2 \Lambda \tphi^{q}\,,
	\ee
to Eq.~(\ref{inteq}). After integrating, we obtain
	\be 
	N_N = \frac{ 2^{p/(1+p)} (1+p)^3 V_0^{p/(1+p)}  }{ q (pq +2p +2)} \[ \tphi_N^{\frac{pq+2p+2}{1+p}} - \tphi_e^{\frac{pq+2p+2}{1+p}} \].
\label{nn:pwl}
	\ee
Using the condition $\epsilon = 1$ at the end of inflation, we can calculate $\tphi_e$ as
	\be
	\tphi_e^{\frac{pq+2p+2}{1+p}} = \frac{ 2^{-\frac{2p+1}{1+p}} q^2 (2p+1)}{  (1+p)^3 V_0^{p/(1+p)} }\,. 
\label{phie:pwl}
\ee
Substituting the above expression into Eq.~(\ref{nn:pwl}), we get
	\be
	N_N + N_* = \frac{ 2^{p/(1+p)} (1+p)^3 V_0^{p/(1+p)} }{ q (pq+2p+2)} \tphi_N^{\frac{pq+2p+2}{1+p}}\,,
\label{nns:pwl}	
\ee
where 
	\be \label{n*:power-law} 
	N_* \equiv \frac{q\(2p+1\)}{2\(pq+2p+2\)}\,.
	\ee
Then we can calculate
	\be
	\epsilon_N = \frac{N_*}{N_N + N_*}\,,
	\quad\mbox{and}\quad
	\eta_N = \left[\frac{q^{2p+2} (1+p)^{2 (pq - p - 1)}}{ 2^{2p}  V_0^{2p} (pq+2p+2)^{2(pq+p+1)}}
(N_N + N_*)^{-2(pq + 1+p)}
\right]^{1/(pq+2p+2)} \,.
\label{epneta:pwl}
	\ee
Using Eqs.~(\ref{fc:slow}) and (\ref{nns:pwl}),
we have
\be \label{fcn:power}
\fc(N) 
= \fc{}_* (N + N_*)^\frac{p q}{pq+2p+2}
= \fc{}_* \(\frac{q (2p + 1)}{2 (pq + 2p + 2) \epsilon}\)^\frac{p q}{pq+2p+2}\,, 
\ee
where
\be
\fc{}_* \equiv
\left(\frac{4^p V_0^{2p} q^{qp} \(pq+2p+2\)^{qp}}{(1+p)^{2qp - 2p - 2}}
\right)^{1/(pq+2p+2)}\,.
\ee

\subsection{Numerical results}
\begin{figure}
  	\includegraphics[width=0.7\linewidth]{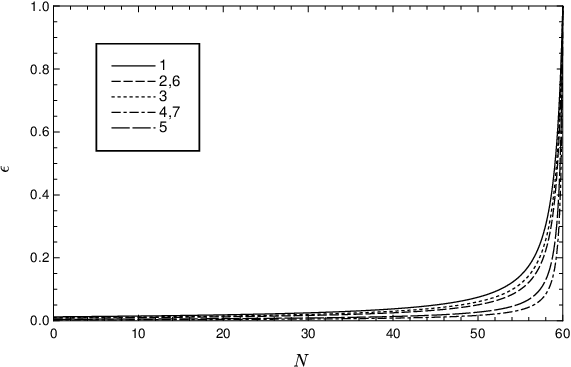}
  	\caption{Plots of slow-roll parameter $\epsilon$ as a function of number of e-folding
for the models 1 - 7. In the plots, models 1 - 7 correspond to lines 1 - 7,  respectively.}
\label{fig1}
	\end{figure}

\begin{figure}
 	\includegraphics[width=0.7\linewidth]{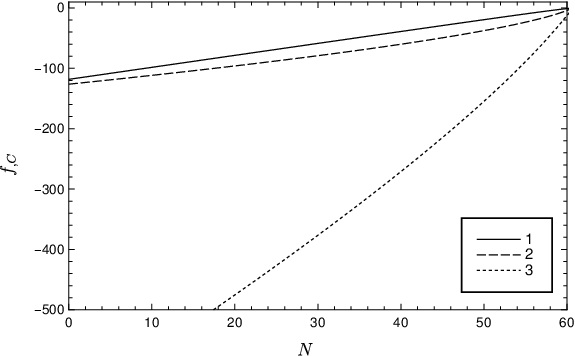}
  	\caption{Plots of $\fc$ as a function of number of e-folding.
In the plots, lines 1 - 3 represent models 1 - 3, respectively.}
  	\label{fig2}
	\end{figure}

\begin{figure}
 		 \includegraphics[width=0.7\linewidth]{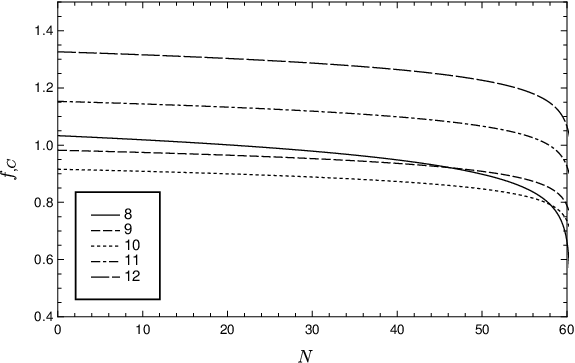}
 		 \caption{Plots of $\fc$ as a function of number of e-folding.
In the plots, lines 8 - 12 represent models 8 - 12, respectively.
}
  	\label{fig3}
	\end{figure}

In this subsection, we solve the evolution equations for the background universe numerically and plot the results in Figs.~(\ref{fig1})--(\ref{fig3}).
The models in our plots are shown in Table~(\ref{table1}).
\begin{table}[]
\begin{tabular}{c|cccccc}
\hline
 No. Model& 1 & 2 & 3 & 4 & 5 & 6  \\
\hline
potential & $e^{\phi}$  & $\phi^{2}$ & $\phi^{4}$ & $0.7 \phi^{1/2}$ & $0.085 \phi^{1/2}$ & $0.002 \phi^{2}$ \\
$p$ & 1 & 1 & 1 & 1 & 1/5 & 1/21	\\
\hline
\hline
 No. Model& 7 & 8 & 9 & 10 & 11 & 12  \\
\hline
potential & $0.05 \phi^{1}$ & $0.02 \phi^{1}$ & $0.0005 \phi^{2}$ & $0.0001 \phi^{2}$ & $0.02 \phi^{2}$ & $0.5 \phi^{2}$		\\
$p$ & 1/10 & 1/5 & 1/21 & 1/21 & 1/21 & 1/21 \\
\hline
\end{tabular}
\caption{Models used in the numerical calculation}
\label{table1}
\end{table}
In Fig.~(\ref{fig1}), we plot evolution of $\epsilon$ for both exponential and power-law potentials cases.
From this figure, we see that  for both forms of potential, inflationary epoch can be taken place such that slow-roll parameter $\epsilon$ increases from small value during early stage towards one at the end of inflation.
The main different feature of the models comes from different evolution of $\fc$.
As will be seen in the next section, $\fc$ controls evolution of the curvature perturbation during inflation.
Evolutions of $\fc$ are plotted in Figs.~(\ref{fig2}) and (\ref{fig3}).
According to Eq.~(\ref{fc:exp}), $\fc$ is proportional to $1/\epsilon$ for the exponential potential, so that for this form of potential $\fc$ can increase several order of magnitude through out inflationary epoch.
This conclusion agrees with the plot in Fig.~(\ref{fig2}). 
However, for the power-law potential, Eq.~(\ref{fcn:power}) shows that the rate of change of $\fc$ decreases when $q$ and $p$ decrease.
When $p \to 0$,  the model for power-law case reduces to Einstein gravity such that $\fc=1$.  
Nevertheless, it follows from Eq.~(\ref{epneta:exp}) that there is no Einstein limit for the case of the exponential potential.
Dependence of $\fc$ on parameters $p$ and $q$ for the case of power-law potential is shown in Figs.~(\ref{fig2}) and (\ref{fig3}).
From the figures, we see that the variation of $\fc$ reduces when $p$ and $q$ decrease.   

\section{Evolution of primordial density perturbations}
\label{sec:4}

In this section we consider evolution of primordial perturbations generated in inflationary model introduced in Sec.~(\ref{sec:2}).
In the following consideration, we concentrate on scalar and tensor perturbation which usually provide predictions of the model,
and set $\mpl = 1$.

\subsection{Scalar perturbations}

To study linear perturbations in this theory, we parametrize perturbation in lapse function and shift vector as
\begin{eqnarray}
\N=a(\tau)\,(1+\alpha)\,,\qquad \N_{i}=a(\tau)\partial_{i}\chi\,,\label{alphachi} 
\end{eqnarray}
where $\alpha$ and $\chi$ are scalar perturbations, $\tau = \int dt/a$ is the conformal time and the background of the lapse function is scale factor $a(\tau)$.
The induce metric can be decomposed in terms of scalar perturbation as
\begin{equation}
h_{ij}=a(\tau)^{2}\,\((1+2\xi)\delta_{ij}+2\,\partial_{i}\partial_{j}E\)\,,\label{Ezeta} 
\end{equation}
where $\xi$ and $E$ are the other scalar-perturbation variables.
Due to the spatial diffeomorphism invariant, we can set spatial gauge degrees of freedom such that $E=0$.
Here, we use the perturbation variables introduced in \cite{Aoki:20}: 
\begin{eqnarray}
\frac{\delta\rho_\phi}{\rho_\phi}\equiv
\delta_\phi + 3 \frac{a'}{a^{2}} (1 + w_\phi)\,\chi\,,
\qquad
\alpha\equiv\Psi -\frac{\chi'}{a}\,,\qquad
\xi\equiv-\Phi-\frac{a'}{a^{2}}\,\chi\,,\qquad 
v_\phi\equiv-\frac{a}{k^{2}}\,\theta_{\phi}+\chi\,,\label{eq:def-thetai}
\end{eqnarray}
where $\Psi$, $\Phi$, $\delta_{\phi}$ and $\theta_{\phi}$ describe the two metric perturbations, the density contrast in the inflaton field and the velocity perturbation in the inflaton field, respectively.
In the above and subsequent expressions, The wave number of the perturbations modes in the Fourier space is denoted by $k$, and a prime denotes derivative with respect to conformal time. 
These perturbation variables can reduce to gauge invariant combinations in Newtonian gauge in the context of Einstein gravity. 
Usually, perturbations during inflation are described by curvature perturbation in the comoving gauge, because the curvature perturbation in this case is constant on large scales when entropy perturbations and isotropic perturbations disappear \cite{Wands:2000dp}.
For this reason, we study the scalar perturbations using the following perturbation variables:
\begin{eqnarray}
\tilde{\delta}_{\phi}\equiv
\delta_{\phi} + \frac{3}{k^2} \frac{a'}{a} \( 1 + w_{\phi} \) \theta_{\phi} \,,
\qquad
\zeta\equiv \Phi + \frac{a'}{a} \frac{\theta_{\phi}}{k^2}\,,\qquad
\bar{\alpha}\equiv \Psi - \frac{a'}{a} \frac{\theta_{\phi}}{k^2} - \frac{\theta'_{\phi}}{k^2} \,.\label{eq:def-comoving}
\end{eqnarray}
The above perturbation variables can become gauge invariant combinations in the comoving gauge for the case of Einstein gravity.

In principle, to investigate the primordial density perturbations generated during inflation, we should construct the action for second order perturbation in which the primordial perturbation are described by canonical variables.
However, the action for perturbation for this theory is rather complicated due to the scale-dependence of gauge fixing term in the action.
Thus instead of constructing this action, we concentrate on evolution of the curvature perturbation on large scales which its evolution equation can be obtained from the evolution equations for perturbations  presented in \cite{Aoki:20}.
The necessary equations are 
\ba
0 &=& 
\Phi'+\cH\,\Psi
+\frac{4k^{2}\cH f_{,CC}}{a^2 f_{,C}}\,\Phi
+\frac{3}{\Gamma_{1}k^{2}}
\left[
6\Gamma_{2}\cH^{2}\frac{f_{,CC}}{a^2 f_{,C}}
-\frac{1}{2}\Gamma a^{2}f_{,C}^{2}
-\frac{k^{2}\Gamma_{1}}{9\Gamma_{2}}\right]
\eta \cH^{2}\,\theta_{\phi}	\,,\label{eq:e1}\\
0 &=& 
-\frac{2}{3}\,f_{,C}\,\frac{k^{2}}{a\cH}\,\Phi
-\frac{a}{3\cH} \rho_{\phi}\,\delta_{\phi}
-\frac{3a\Gamma_{2} \Gamma}{k^{2}\Gamma_{1}}\theta_{\phi} \,,\label{eq:e-psi}\\
0 &=&
\Psi+\frac{(4f_{,CC}k^{2}-a^{2}f_{,C})\Gamma_{1}}{a^{2}\Gamma_{2}}\,\Phi
-\frac{a^{2}(f_{,C}^{2}-1)}{\Gamma_{2}}\,c_e^2\,\delta_{\phi}\nonumber \\
 && +  \frac{1}{k^{2}\Gamma_{1}\Gamma_{2} a^2}\left[a^{4}f_{,C}(f_{,C}^{2}-1)(\Gamma'+2\cH \Gamma)+2\cH \frac{f_{,CC}}{f_{,C}}\left(2f_{,C}^{2}k^{2}\Gamma_{1}-9a^{2}\Gamma\Gamma_{2}\right)\right]\eta \cH^2 \theta_{\phi}\,, \label{eq:e2}
\ea
where $\cH \equiv a'/a$ and 
\ba
\Gamma &\equiv& \frac{1}{3}\(\rho_{\phi}+p_{\phi}\)\,,\\
\Gamma_{1}&\equiv&\Gamma\,a^{2}f_{,C}+\frac{2}{9}k^{2}\,,\\
\Gamma_{2}&\equiv&\Gamma\,a^{2}+\frac{2}{9}k^{2}f_{,C}\,.
\ea
The Eqs.~(\ref{eq:e1})--(\ref{eq:e2}) are completed by conservation equations for the perturbations,
\begin{eqnarray}
0 &=&\delta'_{\phi}
+(1 + w_{\phi})\,\theta_{\phi}
-3 \cH \left(w_\phi - c_e^2\right)
\delta_{\phi}
- 3\(1 + w_\phi\)\,\Phi' \,,\label{Eqmatter1}
\\
0 &=&
\theta'_{\phi}-k^{2}\Psi
-\frac{c_e^2}{1 +w_\phi}\,\,k^{2}\,\delta_{\phi}
+\cH \left(1-3 c_e^2\right)\theta_{\phi}\,.\label{Eqmatter2}
\end{eqnarray}
After straightforward calculation, we obtain evolution equation for $\zeta$ as  
\be \label{standard form}
 \zeta'' + K_{1} \zeta' + K_{2} \zeta = 0 \,,
\ee
where coefficients $K_{1}$ and $K_{2}$ are function of number of e-folding, wavenumber $k$ and $\cH$.
The explicit expression of these coefficients are presented in the appendix.
The curvature perturbation $\zeta$ is related to the perturbation in scalar field as 
\be
\zeta = - \frac{a^2}{2\fc k^2} \rho_{\phi}\,\delta_{\phi} + \( \cH - \frac{3 a^2 \cH \Gamma \Gamma_{2}}{2 k^2 \fc \Gamma_{1}}  \) \frac{\theta_{\phi}}{k^2} \,.
\ee
In the region where $k^2/ \cH^2 > \mathcal{O}(\epsilon) $, Eq.~(\ref{standard form}) can be written up to the lowest order in slow-roll parameters as 
\be
	v'' - \frac{z''}{z}v + c_s^2 k^2 v = 0 \,, 
\label{v1}
\ee
where $v=z \zeta$ and in this case 
\be
	\frac{z''}{z} = \frac{1}{4} (8 + 18 \fc - 9 \fc^2 - 18 \fc^3 + 9 \fc^4) \cH^2 \, ,\\
\quad\mbox{and}\quad 
c_s^2 =1+ \mathcal{O}(\epsilon).
\ee
The expression for $z$ is computed from
\be
z =a\, exp\left\{ \int d\tau \( \frac{3}{2}\fc \cH \( 1 - \fc \) \) \right\}\,, 
\ee
where $z$ reduces to $z = a$ in the Einstein limit.
For the subhorizon modes, $k \gg \cH$, Eq.~(\ref{v1}) is satisfied by the solution \cite{mukhanov:99} 
\be
v = \frac{e^{-i k c_s \tau}}{\sqrt{2 c_s k}}\,.
\label{vsub}
\ee
For the superhorizon modes, where $k\ll \cH$ but $k^2/ \cH^2$ is still larger than  $\mathcal{O}(\epsilon) $,
Eq.~(\ref{v1}) is solved by the solution $v \propto z$, where the proportional constant could be computed by matching the solution for the subhorizon limit with that for the superhorizon limit.
However, we are not interested in such calculation here because the condition $k/ \cH > \mathcal{O}(\epsilon) $ is violated just a few numbers of e-folding after the horizon crossing. 
When this condition is violated, the evolution of $\zeta$ is time dependent as we will see below.
For the case where $k^2/ \cH^2 < \mathcal{O}(\epsilon) $, the evolution equation for the curvature perturbation up to the dominant contribution from $k / \cH$ can be written in  the form
\be \label{eqzeta-wrtN}
\frac{d^2 \zeta_k}{d N^2} + \(3 + A\)\frac{d \zeta_k}{dN}
+ \( \Xi + B \) \zeta_k = 0\,.
\ee
Here,
\ba
A &\equiv&
\frac{1}{18 \eta \( \fc -1  \)^2} \Big[ \eta^2 \fc^2 \big\{ 9 + \fc \( \Xi -9 \) \big\} + 4 \Big\{ \epsilon^2 \( 4 \fc -3 \) - \eta_1 \big( 3 \fc^5 - 6 \fc^4 + 6 \fc^2 + \epsilon \fc \nonumber\\
&&- 6 \fc - \epsilon +3 \big) - 6 \( 3 \fc^5 -6 \fc^4 + 6 \fc^2 - 4 \fc +1 \) + \epsilon \big( 6 \fc^5 -12 \fc^4 + 12 \fc^2 -17 \fc \nonumber\\
&&+ \epsilon_1 \( \fc -1 \) + 8 \big) \Big\}\Xi + 2 \eta \Big\{ 5 \fc^2 \Xi - 2 \fc \Xi - \epsilon \big( 18 \fc^4 - 36 \fc^3 + \fc^2 (5 \Xi -27) \nonumber\\
&&- 3 \fc (\Xi -27) -36 \big) + 27 \fc^4 - 27 \fc^3 + 9 \eta_1 (\fc-1)^3(\fc+1) - 81 \fc^2 + 135 \fc -54 \Big\} \Big]		\,,\\
B &\equiv& 
-\frac{1}{54 \eta^2 (\fc-1)^2}\fc \Big[4 \eta \Xi \Big\{\epsilon \big(-\epsilon_1 (\fc-1) (\fc (\Xi-9)+9)-6 \fc^6 \Xi+12 \fc^5 \Xi+108 \fc^4\nonumber\\
&&-12 \fc^3 (\Xi+18)+\fc^2 (37 \Xi-99)+\fc (387-19 \Xi)-180\big)+\eta_1(\fc-1) \big(\epsilon \big(9 \fc^3-9 \fc^2 \nonumber\\
&&+\fc (\Xi-18)+18\big)+3 \left(\fc^5 \Xi-\fc^4 \Xi-\fc^3 (\Xi+9)+\fc^2 (\Xi+9)-\fc (\Xi-18)-18\right)\big) \nonumber\\
&&+\epsilon^2 \left(-18 \fc^4+36 \fc^3-9 \fc^2 (\Xi-4)+\fc (6 \Xi-99)+45\right)+3 \big(6 \fc^6 \Xi-12 \fc^5 \Xi \nonumber\\
&&-\fc^4 (\Xi+54)+2 \fc^3 (7 \Xi+54)+\fc^2 (18-13 \Xi)+2 \fc (\Xi-72)+\Xi+72\big)\Big\} \nonumber\\
&&+8 (\epsilon-3) \Xi^2 \Big\{\epsilon \left(\epsilon_1 (\fc-1)+6 \fc^5-12 \fc^4+12 \fc^2-17 \fc+8\right)+\eta_1  \big(-\epsilon \fc+\epsilon-3 \fc^5  \nonumber\\
&&+6 \fc^4-6 \fc^2+6 \fc-3\big)+\epsilon^2 (4 \fc-3)-6 \left(3 \fc^5-6 \fc^4+6 \fc^2-4 \fc+1\right)\Big\}\nonumber\\
&&-2 \eta^2 \Big\{9 \fc^5 \Xi (\eta_1 -2 \epsilon+6)-18 \fc^4 \Xi (\eta_1 -2 \epsilon+6)+\fc^2 \big(9 \Xi (2 \eta_1 -10 \epsilon+13)  \nonumber\\
&&+(3 \epsilon-2) \Xi^2+243\big)-9 \fc (\Xi (\eta_1 -4 \epsilon+5)+27)+\fc^3 \left((8-6 \epsilon) \Xi^2+18 (2 \epsilon-1) \Xi-81\right) \nonumber\\
&&+81\Big\}+\eta^3 \left(-\fc^2\right) \Xi \left(\fc^2 \Xi-9 \fc+9\right)\Big]
- \Xi \,.
\ea
where $\Xi \equiv k^2 /\cH^2 < {\cal O}(\epsilon) \ll 1$, $\epsilon_1 \equiv \dot\epsilon/(H\epsilon)$ and $\eta_1\equiv \dot\eta/(H\eta)$. 
Since the analytic solution for the above equation  is difficult to be computed due to time dependence of $\fc$, which is not necessary slowly varying with time, we will study the important features of the solution for this equation numerically in the next section.
However, from the structure of this equation,
we expect that the dominant solution for Eq.~(\ref{eqzeta-wrtN}) should be time dependent unless $\fc = 1$.
One can check that for $\fc =1$, coefficients $A$ and $B$ vanish, which corresponds to Einstein gravity. 
This also indicates that $\zeta$ is nearly constant on large scales when $\fc$ is sufficiantly close to unity.

\subsection{Numerical result}

\begin{figure}
  	\includegraphics[width=0.49\linewidth]{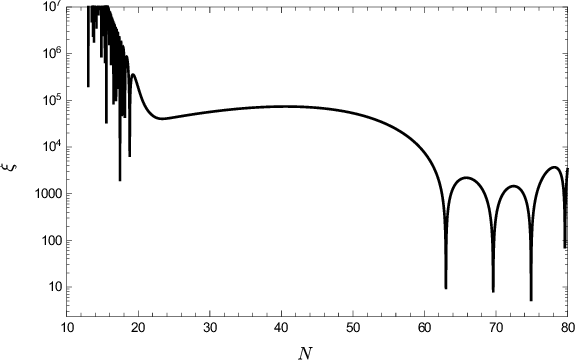}
	\includegraphics[width=0.49\linewidth]{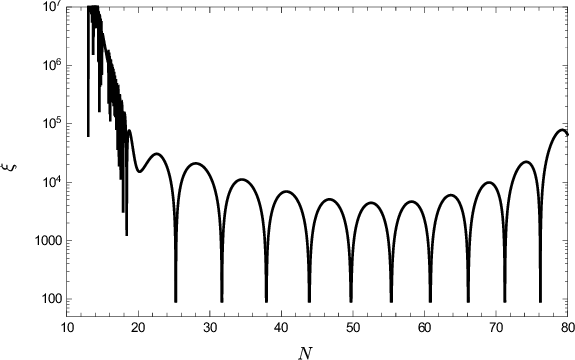}
\\
	\includegraphics[width=0.49\linewidth]{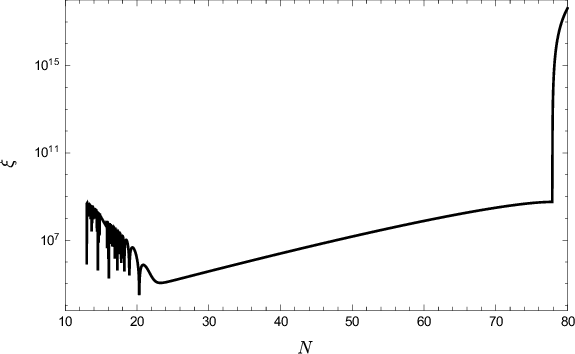}
	\includegraphics[width=0.49\linewidth]{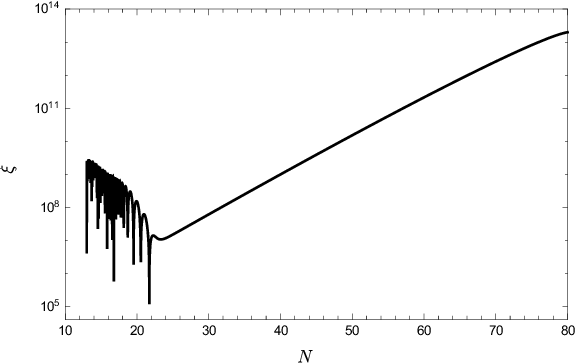}
  	\caption{Plots of $\zeta$ as a function of the number of e-folding.
The left top, right top, left bottom and right bottom panels represent evolutions of $\zeta$ for models 8, 10, 11 and 12, respectively.
In all plots, the perturbation crosses Hubble radius at $N = 20$.
  }
\label{fig:4}
	\end{figure}

To confirm rough analytic estimation in the previous section, we solve the evolution equation for curvature perturbation numerically.
We start the numerical integration at the time when physical wavelength of perturbation is well inside the Hubble radius. 
The initial conditions are chosen according to Eq.~(\ref{vsub}) by spitting $\zeta$ to the real and the imaginary parts.
We integrate Eq.~(\ref{standard form}) for
the real $\zeta_{\rm real}$ and the imaginary $\zeta_{\rm imaginary}$ parts of $\zeta$ separately, and plot the absolute value $\zeta = \sqrt{\zeta_{\rm real}^2 + \zeta_{\rm imaginary}^2}$ in the following figures.
According to discussion in the previous section, the main features of  $\zeta$-evolution depend on $\fc$.
Hence, we consider evolution of $\zeta$ for models 8, 10, 11 and 12 in which $\fc$ varies by a few multiplication factor, $\fc$ is nearly constant with  $\fc \lesssim 1$, $\fc \sim 1$ and $\fc \gtrsim 1$.
From the plots in Fig.~(\ref{fig:4}), we see that $\zeta$ can rapidly grow on super Hubble radius scales although $\fc$ changes only fews percents around one through out inflation.
These results could be consequences of unknown sources of entropy and anisotropic perturbations.
On the other hand, the growth of perturbations on large scales may arise due to the possibility that $\zeta$ is not equivalent to curvature perturbation in the comoving gauge in the standard cosmological perturbation theory.    

\subsection{Tensor perturbations}

To study the tensor modes of perturbation,
we write the metric tensor in the form of the background metric and tensor perturbations as
\be 
h_{ij} = a^2\(\delta_{ij} + \gamma_{ij}\)\,,
\qquad
h^{ij} = a^{-2}\(\delta^{ij} - \gamma^{ij}\)\,,
\label{tensor:met}
\ee
where $\gamma_i^i = 0$ and $\partial_i\gamma^{ij} = 0$.
Since the gauge-fixing term does not depend on tensor quantity, the tensor perturbation does not depend on the gauge and therefore the tensor perturbation computed from Eq.~(\ref{action:ful}) and Eq.~(\ref{action}) are equivalent.
Hence for convenience, we insert the metric from Eq.~(\ref{tensor:met}) into Eq.~(\ref{action}) and expand the action up to second order in perturbation.
We obtain the second order action for the tensor perturbation as
\be
S^{(2)}_T = \int dt dx^3 a^3\(
\frac{1}{8f'} \dot\gamma_{ij}\dot\gamma^{ij}
- \frac{f'}8 \partial_i \gamma^{kl} \partial^i \gamma_{kl}
\)\,,
\label{act:tensor1}
\ee
where the divergent term is omitted.
The tensor perturbation $\gamma_{ij}$ can be expanded in terms of the polarization tensors as
\be
\gamma_{ij} = \int {d^3k \over ( 2 \pi)^3} \sum_{s=\pm} \epsilon_{ij}^s(k)
\gamma^s_{k}(\tau) e^{ i \vec k \cdot \vec x }\,,
\ee
where $\epsilon_{ii} = k^i \epsilon_{ij} =0$ and $\epsilon^s_{ij}(k)
\epsilon^{s'}_{ij}(k)  = 2 \delta_{s s'}$. 
According to the action Eq.~(\ref{act:tensor1}),
each of the mode functions $\gamma^s_{k}(\tau)$ obeys
\be
\gamma^s_k{}^{\prime\prime} + \frac{(a^2 / \fc)'}{a^2 / \fc} \gamma^s_k{}^\prime + k^2 c_T^2 \gamma^s_k = 0\,,
\label{tensor:eom1}
\ee
where $c_T^2 = \fc^2$ is the sound speed squared of the tensor perturbations.
As in the usual calculation,
we define \cite{mukhanov:99}
\be
v^s_T \equiv z_T \gamma^s_k\,,
\quad\mbox{where}\quad
z_T^2 \equiv \frac{a^2 }{4 \fc}\,,
\ee
so that Eq.~(\ref{tensor:eom1}) becomes
\be
v^s_T{}^{\prime\prime} + k^2 c_T^2 v^s_T - \frac{z_T''}{z_T} v^s_T = 0\,.
\label{tensor:eom2}
\ee
Applying the standard calculation,
we have
\be
\left|v^s_T\right|_c^2  = \left.\frac{1}{2 c_T k}\right|_c\,,
\ee
which implies that the amplitude of tensor perturbation is constant on large scale up to the lowest order in slow-roll parameter, and we can compute the power spectrum for the tensor perturbations as
\be
P_{k}^{T}\equiv \frac{k^{3}}{2\pi ^{2}}\(\left|\gamma^+_k\right|_c^{2} + \left|\gamma^-_k\right|_c^{2}\)
= \frac{2}{\pi^2}\frac{H^2}{\fc^2}\,,
  \label{tensor:spec}
\ee
where the tensor perturbations cross the sound horizon at $a H = c_T k$.
The spectral index for the tensor perturbations can be computed as
\be
n_T \equiv \frac{d\ln P_{k}^{T}}{d\ln k}
=  - 2 \epsilon - 2 \frac{\dot{\fc}}{H \fc}\,,
\label{nt}
\ee
where $c_T^2 \equiv \fc^2$.
Using $\dot{c}_s / (H c_s) \simeq \dot{\fc}/ (H \fc)$,
and Eqs.~(\ref{mod:fd}), (\ref{Nfp}) and (\ref{nn:pwl}),
the tensor spectral index for the exponential and the power-law potentials are given by
\be
n_T = - \frac{2}{\( 2p+1 \)} \frac{N_{*}}{N_N + N_{*}} \,,
\ee
where $N_{*}$ for the exponential potential is given by Eq.~(\ref{n*:exponential}) and $N_{*}$ for the power-law potential is given by Eq.~(\ref{n*:power-law}).

\section{Conclusions}
\label{conclude}

We have studied models of inflation in the MMG theory. 
We have concentrated on $f({\bf H})$ gravity in the form $f(C) = -\Lambda (-C/\Lambda)^{1+p}$, where $p$ and $\Lambda$ are constant, while $C$ is given by Eq.~(\ref{constraintC}).
It can be checked that this theory reduces to Einstein gravity when $p=0$.
It is difficult for this theory to drive inflation without introducing an inflaton field.
We have examined inflationary models in which potential of the inflaton takes the exponential and the power-law forms.
We have found that the slow-roll parameter $\epsilon$ is inversely proportional to the number of e-folding similar to case of Einstein gravity.
The expression for $\epsilon$ in the case of power-law potential takes the form as in the Einstein gravity when $p=0$.
Nevertheless, there is no Einstein limit for the case of exponential potential.
According to the evolution equation for the perturbations, it can be seen that evolution equation of perturbations depends on $\fc$.    
For the case of exponential potential, $\fc$ is inversely proportional to $\epsilon$, so that $\fc$ can vary a few order of magnitude through out inflation.
However, for the case of power-law potential, $\fc$ becomes nearly constant when $p$ is close to zero.
From the numerical integration, we have found that the curvature perturbation on large scales can grow extremely large if $\fc$ is significantly vary in time.
The curvature perturbation becomes constant on large scales when $\fc \sim 1$.
It could be possible that the curvature perturbation on large scales is not conserve in this model because the curvature perturbation used here is not equivalent to curvature perturbation in the comoving gauge in standard cosmological perturbation theory. 
On the other hand, non-conservation of curvature perturbation on large scales could be consequences of entropy and anisotropic perturbations. 
In general, it is difficult to define curvature perturbation that is conserved on large scales similar to the curvature perturbation in the comoving gauge in the Einstein theory, because it is not clear whether the entropy and anisotropic perturbations disappear in this $f({\bf H})$ theory.
These questions are left for future investigation. 
For tensor perturbation, the sound speed of tensor mode can significantly deviate from unity and vary with time if $p \neq 0$.

\newpage
\section*{Appendix: The expressions for the coefficients $\alpha$ and $\beta$}
\label{appendix:A}

In this appendix, we present the explicit form of the coefficients $\alpha$ and $\beta$ in Eq.~(\ref{standard form}). 
Firsly, we decompose them as
	\be
	K_{1} = \frac{n_{1}}{d_{1}},\quad	K_{2} = \frac{n_{2}}{d_{2}},
	\ee 
where the expressions of $n_1$, $n_2$, $d_1$ and $d_2$ are given by 
	\ba
	n_1 &=& a_1 + a_2 k_H^2 + a_3 k_H^4 + a_4 k_H^6 + a_5 k_H^8 + a_6 k_H^8 \, ,\nonumber\\
	d_1 &=& 4 (3 \eta + 2 \fc k_H^2) \Bigl(-8 (-3 + \epsilon) \fc^2 k_H^6 + 9 \eta^3 \bigl(-9 + \fc (9 + k_H^2)\bigr) \nonumber\\
&&+ 4 \eta \fc k_H^4 \bigl(18 - 6 \epsilon - 9 \fc + \fc^2 (9 + k_H^2)\bigr) + 6 \eta^2 k_H^2 \bigl(9 - 3 \epsilon - 18 \fc + k_H^2 + \fc^2 (18 + k_H^2)\bigr)\Bigr) \, , \nonumber\\
	n_2 &=& b_1 + b_2 k_H^2 + b_3 k_H^4 + b_4 k_H^6 + b_5 k_H^8 + b_6 k_H^{10} + b_7 k_H^{12} \, ,\nonumber\\
	d_2 &=&6 (3 \eta + 2 \fc k_H^2)^2 \Bigl(-8 (-3 + \epsilon) \fc^2 k_H^6 + 9 \eta^3 \bigl(-9 + \fc (9 + k_H^2)\bigr) \nonumber\\
&&+ 4 \eta \fc k_H^4 \bigl(18 - 6 \epsilon - 9 \fc + \fc^2 (9 + k_H^2)\bigr) + 6 \eta^2 k_H^2 \bigl(9 - 3 \epsilon - 18 \fc + k_H^2 + \fc^2 (18 + k_H^2)\bigr)\Bigr) \, . \nonumber
	\ea
Here, $a_{i}$ are 
	\ba
	a_1 &=& -96 \fc^3 (-2 - 3 \fc + 3 \fc^2) \cH \, ,\nonumber\\
	a_2 &=& -432 \eta \fc^3 (-3 + 2 \fc - 2 \fc^2 + \fc^3) \cH + 16 \fc^3 \bigl(6 \eta_1 + (2 \epsilon -  \eta \fc) (-2 - 3 \fc + 3 \fc^2)\bigr) \cH \, ,\nonumber\\
	a_3 &=& 16 \epsilon \fc^3 (2 \epsilon_1 - 2 \eta_1 + 2 \epsilon -  \eta \fc) \cH - 648 \eta^2 \fc (2 - 5 \fc + 4 \fc^2 - 4 \fc^3 + \fc^4) \cH \nonumber\\
&&+ 72 \eta \fc^2 \bigl(2 \eta_1 (2 + \fc^2) + 2 \epsilon (2 - 5 \fc + 2 \fc^2) -  \eta (-1 + \fc - 2 \fc^2 + \fc^3)\bigr) \cH \, ,\nonumber\\
	a_4 &=& 324 \eta^3 (-4 + 15 \fc - 18 \fc^2 + 6 \fc^3 + 3 \fc^4) \cH + 24 \eta \fc \Bigl(\fc \bigl(6 \epsilon_1 \epsilon + \eta^2 -  \epsilon \eta \fc (3 + 2 \fc^2) \nonumber\\
&&+ \epsilon^2 (2 + 4 \fc^2)\bigr) + \eta_1 \bigl(-2 \epsilon \fc (2 + \fc^2) + \eta (-1 + \fc^4)\bigr)\Bigr) \cH - 36 \eta^2 \Bigl(\eta (2 - 3 \fc + \fc^2 \nonumber\\
&&- 6 \fc^3 + 9 \fc^4) - 6 \fc \bigl(\epsilon (6 - 11 \fc + 6 \fc^2 + 2 \fc^3 - 2 \fc^4) + \eta_1 (1 + \fc + \fc^2 -  \fc^3 + \fc^4)\bigr)\Bigr) \cH \, ,\nonumber\\
	a_5 &=& 36 \eta^2 \fc \Bigl(6 \epsilon_1 \epsilon - 2 \epsilon^2 - 3 \epsilon \eta \fc + 8 \epsilon^2 \fc^2 + \eta^2 \fc^2 - 2 \epsilon \eta \fc^3 + \eta_1 \bigl(\eta \fc (-1 + \fc^2) \nonumber\\
&&- 2 \epsilon (1 + 2 \fc^2)\bigr)\Bigr) \cH - 54 \eta^3 \Bigl(\epsilon (-20 + 66 \fc - 60 \fc^2 - 24 \fc^3 + 24 \fc^4) + \fc \bigl(\eta (4 - 3 \fc + 6 \fc^2) \nonumber\\
&&- 6 \eta_1 (2 -  \fc - 2 \fc^2 + 2 \fc^3)\bigr)\Bigr) \cH \, ,\nonumber\\
	a_6 &=& 54 \eta^3 \bigl(2 \epsilon_1 \epsilon + \eta_1 \eta \fc (-1 + \fc^2) + \epsilon^2 (-2 + 4 \fc^2) + \epsilon \fc (\eta - 2 \eta_1 \fc - 2 \eta \fc^2)\bigr) \cH \, ,\nonumber\\
	\ea
and $b_{i}$ are
	\ba
	b_1 &=& 576 \fc^4 \cH^2 + 5184 \fc^4 (-2 + \fc + \fc^2) \cH^2 \, ,\nonumber\\
	b_2 &=& 192 \fc^4 (-2 \epsilon + \eta \fc) \cH^2 + 2592 \eta \fc^3 (-18 + 18 \fc -  \fc^2 - 4 \fc^3 + 5 \fc^4) \cH^2 \nonumber\\
&&+ 864 \fc^3 \Bigl(\eta (1 -  \fc + \fc^3 + 2 \fc^4) - 2 \fc \bigl(\eta_1 -  \eta_1 \fc + \epsilon (-6 + 3 \fc + 2 \fc^2)\bigr)\Bigr) \cH^2 \, ,\nonumber\\
	b_3 &=& 16 \fc^4 (-2 \epsilon + \eta \fc)^2 \cH^2 + 7776 \eta^2 \fc^2 (-9 + 18 \fc - 11 \fc^2 - 5 \fc^3 + 7 \fc^4) \cH^2 \nonumber\\
&&+ 144 \fc^2 \Bigl(\eta^2 (1 + 4 \fc^2 -  \fc^3 + 4 \fc^4 + \fc^6) - 4 \epsilon \fc^2 \bigl(\epsilon_1 (-1 + \fc) \fc + \epsilon (6 - 3 \fc - 2 \fc^2) \nonumber\\
&&+ \eta_1 (-2 + \fc + \fc^2)\bigr) + 2 \eta \fc \bigl(\epsilon (-3 + \fc - 3 \fc^3 - 3 \fc^4) + 2 \eta_1 (-1 + \fc^4)\bigr)\Bigr) \cH^2 \nonumber\\
&&+ 432 \eta \fc^2 \Bigl(\eta (-5 - 4 \fc - 3 \fc^2 + 10 \fc^3 + 8 \fc^4 - 6 \fc^5 + 6 \fc^6) + 6 \fc \bigl(2 \epsilon (9 - 8 \fc + \fc^2 \nonumber\\
&&+ 2 \fc^3 - 2 \fc^4) + \eta_1 (-3 + 5 \fc - 2 \fc^2 -  \fc^3 + \fc^4)\bigr)\Bigr) \cH^2 \, ,\nonumber\\
	b_4 &=& 34992 \eta^3 (-1 + \fc)^2 \fc (-1 + 3 \fc + 2 \fc^2) \cH^2 + 648 \eta^2 \fc \Bigl(\eta (-5 + \fc - 14 \fc^2 + 17 \fc^3 - 5 \fc^4 \nonumber\\
&&- 18 \fc^5 + 18 \fc^6) + 18 \fc \bigl(\eta_1 (-1 + \fc)^2 (-1 + \fc + \fc^2) + 2 \epsilon (3 - 5 \fc + 3 \fc^2 + 2 \fc^3 - 2 \fc^4)\bigr)\Bigr) \cH^2 \nonumber\\
&&+ 72 \eta \fc \biggl(\eta^2 (-1 + 12 \fc^2 - 6 \fc^3 + 31 \fc^4 - 6 \fc^5 + 6 \fc^6) + 2 \eta \fc \bigl(2 \epsilon (1 + 3 \fc + 2 \fc^2 - 18 \fc^3 + 3 \fc^5 \nonumber\\
&&- 3 \fc^6) + 3 \eta_1 (-3 - 2 \fc^2 + \fc^3 + 4 \fc^4 -  \fc^5 + \fc^6)\bigr) - 12 \epsilon \fc^2 \Bigl(\eta_1 (-6 + 5 \fc + \fc^2 -  \fc^3 + \fc^4) \nonumber\\
&& + 2 \bigl(2 \epsilon_1 (-1 + \fc) \fc + \epsilon (9 - 7 \fc + \fc^3 -  \fc^4)\bigr)\Bigr)\biggr) \cH^2 - 24 \fc^2 \biggl(16 \epsilon^3 \fc^2 (-1 + \fc^2) \nonumber\\
&&-  \eta^2 \fc (1 + \fc^2) \bigl(\eta + 2 \eta \fc^2 + 2 \eta_1 \fc (-1 + \fc^2)\bigr) - 4 \epsilon^2 \fc \bigl(2 \eta_1 \fc (-1 + \fc^2) + \eta (1 + \fc^2 + 4 \fc^4)\bigr) \nonumber\\
&& + 2 \epsilon \eta \Bigl(\eta (1 + \fc^2)^2 (1 + 2 \fc^2) + 2 \fc \bigl(\epsilon_1 -  \epsilon_1 \fc^2 + 2 \eta_1 (-1 + \fc^4)\bigr)\Bigr)\biggr) \cH^2 \, ,\nonumber\\
	b_5 &=& 8748 \eta^4 \fc (9 - 13 \fc + 2 \fc^2 + 2 \fc^3) \cH^2 + 972 \eta^3 \fc \bigl(6 \eta_1 (-1 + 7 \fc - 6 \fc^2 - 3 \fc^3 + 3 \fc^4) \nonumber\\
&&- 12 \epsilon (-3 + 12 \fc - 11 \fc^2 - 6 \fc^3 + 6 \fc^4) + \eta (2 - 11 \fc + 16 \fc^2 - 16 \fc^3 - 18 \fc^4 + 18 \fc^5)\bigr) \cH^2 \nonumber\\
&&+ 36 \eta \fc^2 \Bigl(-4 \epsilon^2 \eta + 48 \epsilon^3 \fc - 8 \epsilon \eta^2 \fc + 4 \epsilon^2 \eta \fc^2 + 5 \eta^3 \fc^2 - 48 \epsilon^3 \fc^3 - 24 \epsilon \eta^2 \fc^3 + 40 \epsilon^2 \eta \fc^4 \nonumber\\
&&+ 5 \eta^3 \fc^4 - 8 \epsilon \eta^2 \fc^5 + 8 \epsilon_1 \epsilon \eta (-1 + \fc^2) + 2 \eta_1 (-1 + \fc^2) \bigl(12 \epsilon^2 \fc + \eta^2 \fc (1 + 2 \fc^2) \nonumber\\
&&- 2 \epsilon \eta (3 + 5 \fc^2)\bigr)\Bigr) \cH^2 + 108 \eta^2 \fc \Bigl(72 \epsilon^2 \fc (-3 + 4 \fc - 2 \fc^2 -  \fc^3 + \fc^4) + \eta^2 \fc (-2 - 3 \fc \nonumber\\
&&+ 23 \fc^2 - 15 \fc^3 + 9 \fc^4) + 6 \eta_1 (-1 + \fc) \bigl(-6 \epsilon \fc (2 -  \fc + \fc^3) + \eta (1 + \fc + 5 \fc^2 + 2 \fc^3 + 3 \fc^5)\bigr) \nonumber\\
&&+ \epsilon \bigl(-72 \epsilon_1 (-1 + \fc) \fc^2 + \eta (10 + 6 \fc + 44 \fc^2 - 114 \fc^3 + 66 \fc^4 + 36 \fc^5 - 36 \fc^6)\bigr)\Bigr) \cH^2 \, ,\nonumber\\
	b_6 &=& 4374 \eta^4 \fc \bigl(2 \eta_1 (2 - 2 \fc -  \fc^2 + \fc^3) + \eta \fc (2 - 3 \fc - 2 \fc^2 + 2 \fc^3) - 2 \epsilon (7 - 8 \fc - 4 \fc^2 + 4 \fc^3)\bigr) \cH^2 \nonumber\\
&&+ 54 \eta^2 \fc \Bigl(-4 \epsilon^2 \eta + 48 \epsilon^3 \fc - 12 \epsilon^2 \eta \fc^2 + \eta^3 \fc^2 - 48 \epsilon^3 \fc^3 - 16 \epsilon \eta^2 \fc^3 + 40 \epsilon^2 \eta \fc^4 + 5 \eta^3 \fc^4 - 8 \epsilon \eta^2 \fc^5 \nonumber\\
&&+ 4 \epsilon_1 \epsilon \eta (-1 + \fc^2) + 4 \eta_1 (-1 + \fc^2) \bigl(6 \epsilon^2 \fc + \eta^2 \fc^3 -  \epsilon (\eta + 5 \eta \fc^2)\bigr)\Bigr) \cH^2 - 972 \eta^3 \fc \Bigl(- \eta^2 (-2 + \fc) \fc^3 \nonumber\\
&&- 4 \epsilon^2 (-3 + 9 \fc - 8 \fc^2 - 3 \fc^3 + 3 \fc^4) + \eta_1 (-1 + \fc) \bigl(\eta \fc (-2 + \fc - 3 \fc^3) + 2 \epsilon (2 - 5 \fc + 3 \fc^3)\bigr) \nonumber\\
&& + 2 \epsilon \fc \bigl(4 \epsilon_1 (-1 + \fc) + \eta (-3 + 7 \fc - 7 \fc^2 - 3 \fc^3 + 3 \fc^4)\bigr)\Bigr) \cH^2 \, ,\nonumber\\
	b_7 &=& 81 \eta^3 \fc (-2 \epsilon + \eta \fc)^2 \bigl(\eta \fc + 2 \eta_1 (-1 + \fc^2) - 4 \epsilon (-1 + \fc^2)\bigr) \cH^2 - 729 \eta^4 \fc \Bigl(-20 \epsilon^2 + 4 \epsilon_1 \epsilon (-1 + \fc) \nonumber\\
&&+ 24 \epsilon^2 \fc + 8 \epsilon \eta \fc + 8 \epsilon^2 \fc^2 - 12 \epsilon \eta \fc^2 + \eta^2 \fc^2 - 8 \epsilon^2 \fc^3 - 4 \epsilon \eta \fc^3 + 4 \epsilon \eta \fc^4 + 2 \eta_1 (-1 + \fc) \bigl(\eta \fc -  \eta \fc^3 \nonumber\\
&&+ 2 \epsilon (-2 + \fc^2)\bigr)\Bigr) \cH^2 \,.\nonumber
	\ea
In the above expressions, $k_H = k/\cH$, and we use Eqs.~(\ref{enconv2}), (\ref{ceqbk}) and (\ref{ep:gen}) to write $\fcc$ as
\be
\fcc = - \frac{1}{12 X} \fc^2 \( \frac{\epsilon - \fc \eta}{2} \) \,.
\ee

\subsection*{Acknowledgement}

JS was supported by Development and Promotion of Science and Technology Talents Project (DPST)  scholarship for his MSc study.


\begin{thebibliography}{1000}

\bibitem{Guth:1980zm}
  A.~H.~Guth,
  Phys.\ Rev.\  D {\bf 23} (1981) 347.
\bibitem{Linde:1981mu}
  A.~D.~Linde,
  Phys.\ Lett.\  B {\bf 108} (1982) 389.
\bibitem{Albrecht:1982wi}
  A.~Albrecht and P.~J.~Steinhardt,
  Phys.\ Rev.\ Lett.\  {\bf 48} (1982) 1220.
\bibitem{Fujii2009}
Y.~Fujii and K.-i~Maeda,
(Cambridge: Cambridge University Press), 2007
\bibitem{Clifton:11}
T.~Clifton, P.~G.~Ferreira, A.~Padilla, and C.~Skordis, 
Physics Reports {\bf 513}, (2012) 1,
[arXiv:1106.2476 [astro-ph.CO]].
\bibitem{cuscut}
N.~Afshordi, D.~J.~H.~Chung, M.~Doran, and G.~Geshnizjani,
Phys.\ Rev.\  D {\bf 75} (2007) 123509,
[astro-ph/0702002].
\bibitem{Iyonaga:18}
A.~Iyonaga, K.~Takahashi, and T.~Kobayashi,
JCAP {\bf 12} (2018) 002,
[arXiv:1809.10935 [gr-qc]].
\bibitem{Iyonaga:20}
A.~Iyonaga, K.~Takahashi, and T.~Kobayashi,
JCAP {\bf 07} (2020) 004,
[arXiv:2003.01934 [gr-qc]].
\bibitem{cov}
J.~Khoury, G.~E.~J.~Miller, and A.~J.~Tolley,
Phys.\ Rev.\  D {\bf 85} (2012) 084002,
[arXiv:1108.1397 [hep-th]].
\bibitem{Mukohyama:17}
C.~Lin and S.~Mukohyama,
JCAP {\bf 10} (2017) 033,
[arXiv:1708.03757 [gr-qc]].
\bibitem{Katsuki:18}
K.~Aoki, C.~Lin, and S.~Mukohyama,
Phys.\ Rev.\  D {\bf 98} (2018) 044022,
[arXiv:1804.03902 [gr-qc]].
\bibitem{Aoki:2018brq}
K.~Aoki, A.~ De~Felice, C.~Lin, S.~Mukohyama, and M.~Oliosi,
JCAP {\bf 01} (2019) 017,
[arXiv:1810.01047 [gr-qc]].
\bibitem{Mukohyama:19}
S.~Mukohyama and K.~Noui,
JCAP {\bf 07} (2019) 049,
[arXiv:1905.02000 [gr-qc]].
\bibitem{DeFelice:2020eju}
A.~De~Felice, A.~Doll, and S.~Mukohyama,
JCAP {\bf 09} (2020) 034,
[arXiv:2004.12549 [gr-qc]].
\bibitem{Gao:20}
X.~Gao and Z.-B.~Yao,
Phys.\ Rev.\  D {\bf 101} (2020) 064018,
[arXiv:1910.13995 [gr-qc]].
\bibitem{Yao:21}
Z.-B.~Yao, M.~Oliosi, X.~Gao and S.~Mukohyama,
Phys.\ Rev.\  D {\bf 103} (2021) 024032,
[arXiv:2011.00805 [gr-qc]].
\bibitem{Hu:21}
Y.-M.~Hu and X.~Gao, 
[arXiv:2103.11463 [astro-ph.CO]].
\bibitem{Gao:14}
X.~Gao,
Phys.\ Rev.\  D {\bf 90} (2014) 081501,
[arXiv:1406.0822 [gr-qc]].
\bibitem{Carballo-Rubio:2018czn}
R.~Carballo-Rubio, F.~Di~Filippo, and S.~Liberati,
JCAP {\bf 06} (2018) 026,
[arXiv:1802.02537 [gr-qc]].
\bibitem{Aoki:20}
K.~Aoki, A.~De~Felice, S.~Mukohyama, K.~Noui, M.~Oliosi, and M.~C.~Pookkillath,
Eur.Phys.J.C {\bf 80} (2020) 708,
[arXiv:2005.13972 [astro-ph.CO]].
\bibitem{Lin:18}
C.~Lin,
JCAP {\bf 05} (2019) 037,
[arXiv:1811.02467 [gr-qc]].
\bibitem{Lin:19}
C.~Lin and Z.~Lalak,
[arXiv:1911.12026 [gr-qc]].
\bibitem{Wands:2000dp}
D.~Wands, K.~A.~Malik, D.~H.~Lyth and A.~R.~Liddle,
Phys.\ Rev.\  D {\bf 62} (2000) 043527,
[arXiv:astro-ph/0003278].
\bibitem{mukhanov:99}
J.~Garriga and V.~F.~Mukhanov,
Phys.\ Rev.\ Lett.B\  {\bf 485} (1982) 219.














\end{thebibliography}
\end{document}